\documentclass[conference]{IEEEtran}
\IEEEoverridecommandlockouts
\usepackage{cite}
\usepackage{amsmath,amssymb,amsfonts}
\usepackage{algorithmic}
\usepackage{graphicx}
\usepackage{textcomp}
\usepackage{xcolor}
\def\BibTeX{{\rm B\kern-.05em{\sc i\kern-.025em b}\kern-.08em
    T\kern-.1667em\lower.7ex\hbox{E}\kern-.125emX}}

\usepackage{hyperref}
\hypersetup{
colorlinks=true,
linkcolor=blue,
filecolor=magenta,
urlcolor=cyan,
citecolor=black,
pdfpagemode=FullScreen
}
\begin{document}

\title{Privacy-preserving and Trusted Threat Intelligence Sharing using Distributed Ledgers}
\author{\IEEEauthorblockN{%
Hisham Ali\IEEEauthorrefmark{1}, Pavlos Papadopoulos\IEEEauthorrefmark{1}, Jawad Ahmad\IEEEauthorrefmark{1}, Nikolaos Pitropakis\IEEEauthorrefmark{1}, Zakwan Jaroucheh\IEEEauthorrefmark{1} and William J. Buchanan\IEEEauthorrefmark{1}
}\\
\IEEEauthorblockA{\IEEEauthorrefmark{1} Blockpass ID Lab, Edinburgh Napier  University, Edinburgh, UK.}
}

\maketitle

\begin{abstract}

Threat information sharing is considered as one of the proactive defensive approaches for enhancing the overall security of trusted partners. Trusted partner organizations can provide access to past and current cybersecurity threats for reducing the risk of a potential cyberattack—the requirements for threat information sharing range from simplistic sharing of documents to threat intelligence sharing. Therefore, the storage and sharing of highly sensitive threat information raises considerable concerns regarding constructing a secure, trusted threat information exchange infrastructure. Establishing a trusted ecosystem for threat sharing will promote the validity, security, anonymity, scalability, latency efficiency, and traceability of the stored information that protects it from unauthorized disclosure. This paper proposes a system that ensures the security principles mentioned above by utilizing a distributed ledger technology that provides secure decentralized operations through smart contracts and provides a privacy-preserving ecosystem for threat information storage and sharing regarding the MITRE ATT\&CK framework.

\end{abstract} 


\IEEEpeerreviewmaketitle

\section{\large Introduction}

Organisations can not afford to defend themselves isolated from the threat landscape due to the emergence of new cyber threats and hence, threat information sharing is considered an indispensable cybersecurity domain. An organisation that has faced a specific range of cyberattacks can help other organizations against adversary attacks of the same type. Threat information sharing is an essential aspect of the cybersecurity domain regarding the protection of organizations or individuals against adversary attacks. Threat information sharing involves the processes of the collection, analysis and sharing of cyber threat information among multiple organizations \cite{bandara2021blockchain}. It ranges from public to private sharing, such as threat intelligence sharing, regionally or globally. Therefore, finding a secure and trusted way to share threat information is crucial to ensure the privacy and reliability of the participant parties. Regardless of the data type, purpose and role, whether public or private, it should traverse through a secure and trusted infrastructure, which could be challenging and costly \cite{jaiman2021user}. Thus, providing a privacy-preserving method to avoid risk is crucial to maintaining confidentiality, availability, and integrity, such as preventing data loss or damage, unauthorized disclosure, data unavailability, or unauthorized alteration. 

Traditional systems for information sharing are commonly centralized networking systems that are controlled by third parties via the internet using location-based addressing (URLs) to reach resources. As a result, data is vulnerable to Denial-of-Service (DoS) attacks due to the possibility of a single point of failure taking place, which is the main drawback of centralized systems \cite{ZHENG2020475}. In contrast, decentralized peer-to-peer networks, which is our solution's core concept, represent a practical and efficient approach for data storage and sharing to combat this issue. Notably, utilizing the Hyperledger Fabric permissioned blockchain framework fortifies the system from a single point of failure and DoS attacks due to its decentralized nature. Additionally, threat information sharing is enabled in a distributed manner, which is being transmitted securely through a reliable, scalable, and performance-efficient infrastructure. Hyperledger Fabric's private and permissioned nature provides flexible identity systems, smart contracts, and unique access control policies through private data  \cite{stamatellis2020privacy}. It ensures that private information stored in the ledger cannot be accessed by unauthorized individuals and only by authorized participants using their authorized identity certificates \cite{iftekhar2021hyperledger}.
   
\begin{figure}[ht!]
\centering
\includegraphics[width=1.03\linewidth]{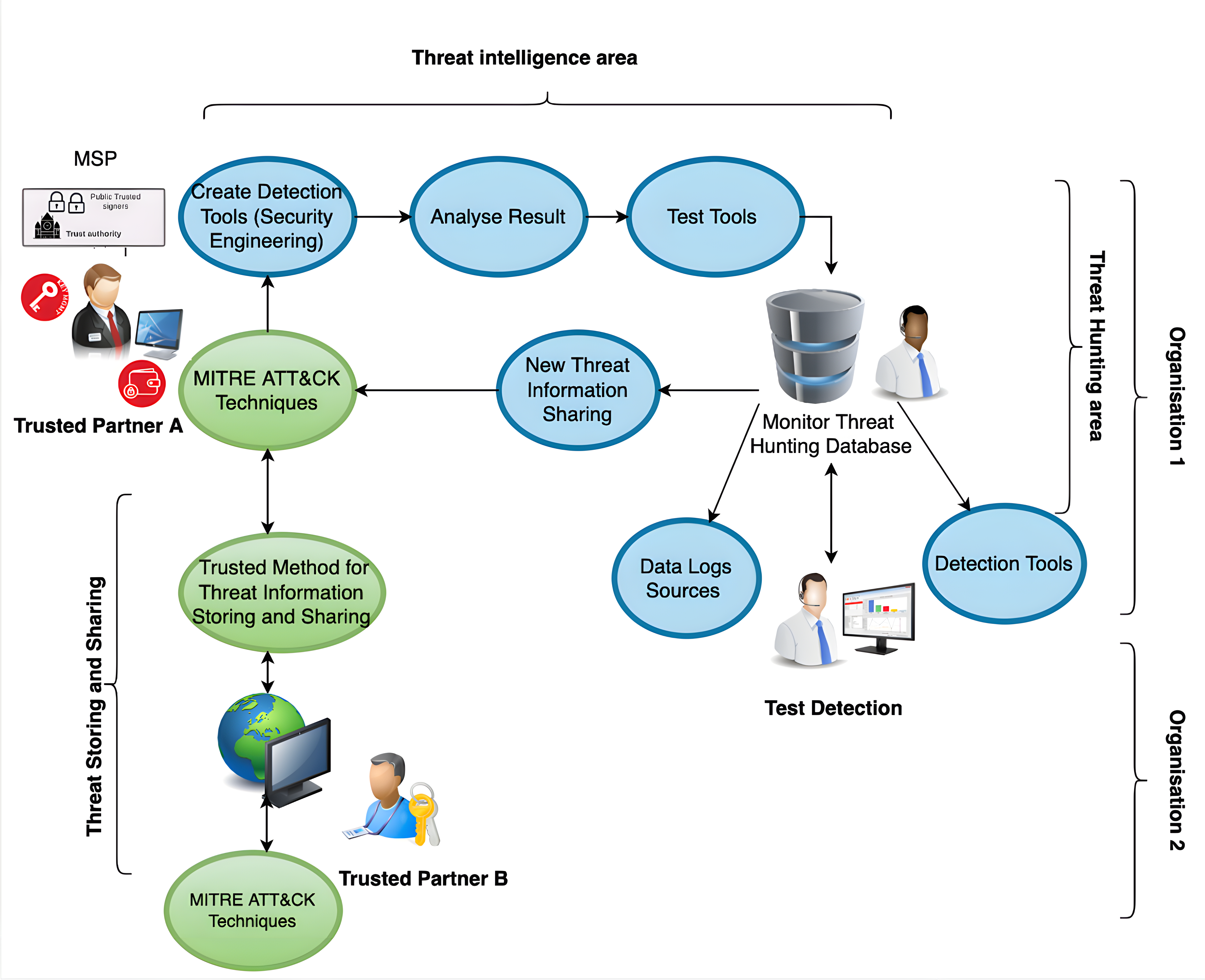}
\caption{MITRE ATT\&CK Framework for Threat Hunting and Sharing.}
\label{fig:mitresharing}
\end{figure}

 We analyze the overlapping between the threat information's tactics and techniques to demonstrate how to use the MITRE Adversarial Tactics, Techniques, and Common Knowledge (ATT\&CK) framework for threat hunting, modelling the adversaries' behaviour and incident response using detection tools and data logs. One can see from Figure~\ref{fig:mitresharing} depicts the MITRE ATT\&CK framework regarding the threat hunting and information sharing. Additionally, the data logs are imported to the MITRE ATT\&CK navigator to review and compare the attacker's behaviors. Furthermore, we present additional data sharing and threat reporting scenarios to look at scaling into other threat reporting areas. In this context, we analyze the use case that has been detected in the healthcare Wicked Panda and Fox Kitten datasets. 

Consequently, in this paper, we propose an integrated distributed infrastructure that combines various privacy-preserving techniques that addresses the security and privacy issues with higher efficiency. The contributions of our work can be summarized as follows:

\begin{itemize}

    \item We analyse, review and compare a range of healthcare IT infrastructures, and then we find the overlapping between tactics and techniques using MITRE ATT\&CK navigator.
    
    \item We propose a trusted ecosystem for threat sharing based on Hyperledger Fabric, Decentralised  Identities and InterPlanetary File System, a distributed storage method.
    
    \item We empirically evaluate the proposed solution in the scope of collaborative operation.
    
\end{itemize}

This paper is organized as follows: \textit{Section 2} presents the literature review and background knowledge. The proposed methodology for threat sharing is discussed in \textit{Section 3}, which includes Data Sources, Data Analysis Framework, Use Case, the Proposed Method and Collaborative Operations using IPFS, Decentralised Identities DIDs and Hyperledger Fabric. \textit{Section 4} presents a security evaluation, discusses potential challenges and issues that need to be carefully considered, and then debates system anonymity consideration, latency, scalability, and throughput. Finally, \textit{Section 5} concludes this work. Few future directions are also outlined in the conclusion section.

\section{\large Literature Review and Background Knowledge}

This section presents the literature review and background information. Several approaches, tools and technologies are integrated into the proposed framework, including MITRE ATT\&CK  IPFS system and Hyperledger Fabric to create privacy-preserving method and trusted information sharing.

\subsection{\large Data Sources and Threat intelligence Sharing}

A wide variety of file sharing methods exist, and they differ from one another in that they allow users to control what, how, and with whom to share. Various studies have been conducted to investigate these properties \cite{alsowailinsider}. The requirements for information sharing can range from the simple sharing of documents to threat intelligence sharing. Threat intelligence is the process of gathering, processing and disseminating information about threats and attackers between trusted partners. Threat intelligence sharing aims to contextualize the information and deliver actionable information that contributes to the decision-making process.
 
\subsection{\large MITRE ATT\&CK Framework}

\begin{figure}[ht!]
\centering
\includegraphics[width=0.9\linewidth]{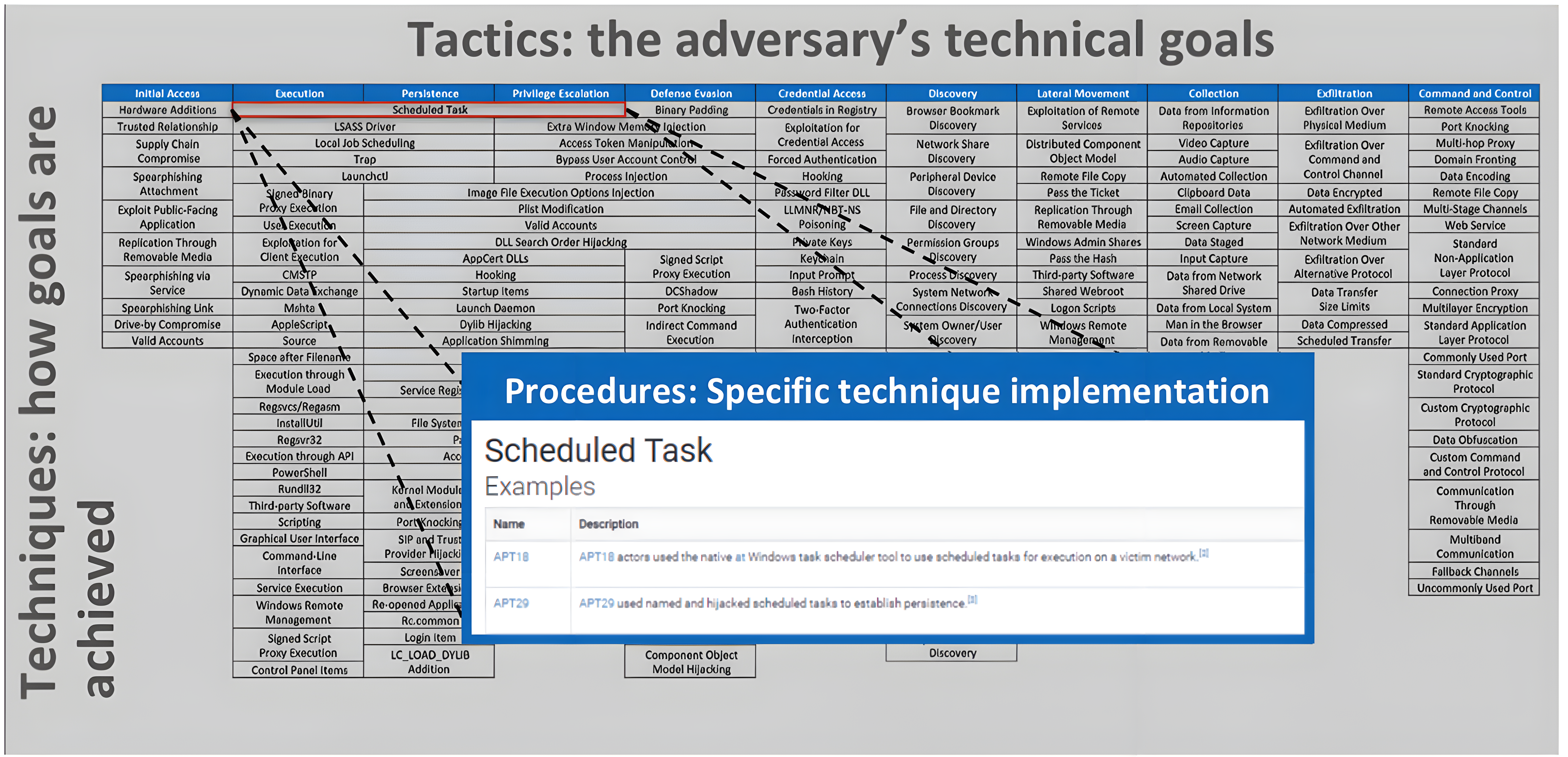}
\caption{Adversary's Tactics and Techniques on MITRE ATT\&CK Framework.  \cite{MITREATTCKNavigatorv2.2}.}  
\label{fig:mitrenavigator}
\end{figure}

The MITRE ATT\&CK framework is an open-source that represents a cumulative knowledge-based on cyber attacker tactics and techniques spotted from previous experiments of the adversary’s behavior \cite{georgiadou2021assessing}. As an essential knowledge base, MITRE ATT\&CK permits the cyber defense teams to review and contrast attacker activity and then understand the best options for defense. It describes the adversary behavior based on actual observations of cybersecurity incidents. MITRE ATT\&CK is used as a navigator to map adversary tactics, techniques, and procedures (TTPS) to its knowledge base as shown in Figure~\ref{fig:mitrenavigator} with their description presented, as follows:

\begin{itemize}
    \item Tactics: the goals the attacker is trying to achieve (the column’s title)
    \item Techniques: the different ways that cyber attackers can achieve the goal of the tactic (the details under the column)
    \item Sub-techniques: data that provides a more detailed explanation of the attacker's technique
    \item Procedures: Examples of malware practices used by cyber attackers
\end{itemize} 

The key notion of using the MITRE ATT\&CK framework is to transform the attacker’s strategies (tactics, techniques) included in the data logs (new threats information) to IDs, scores and colours, respectively for easier conceptualizing and understanding. The MITRE ATT\&CK navigator is used to create a new layer of the threats scenario as shown in Figures~\ref{fig:WICKEDPANDA} and \ref{fig:FoxKitten}. Display and review the existing layer by uploading data logs (JSON) to the navigator, which will be opened in a layer that shows its threat information included in the file, as shown in Figure~\ref{fig:mitrenavigator}. Integrate two layers of different threats in one layer in order to analyse them and find the overlapping threats (tactics and techniques) (green) between them (red and yellow), as shown in Figure~\ref{fig:overlapping}.

We summarise that the MITRE ATT\&CK framework is used to navigate, review and contrast the strategies of attackers, tactics, and techniques to achieve their goals, and find suitable mitigation and countermeasures strategies. Concerning the use case design, MITRE ATT\& CK framework can be used to model the adversary's behaviour after being detected (data logs source). Then the file is encrypted and uploaded to the IPFS, being stored there. Thus, we download its CID  (hash key), which will be sent through Hyperledger Fabric for trusted sharing of threats between the authorized partners. For example, Figure~\ref {fig:mitresharing} illustrates threat hunting and information sharing using the MITRE ATT\&CK framework.

\subsection{\large InterPlanetary File System}

\begin{figure}[ht!]
\centering
\includegraphics[width=1.07\linewidth]{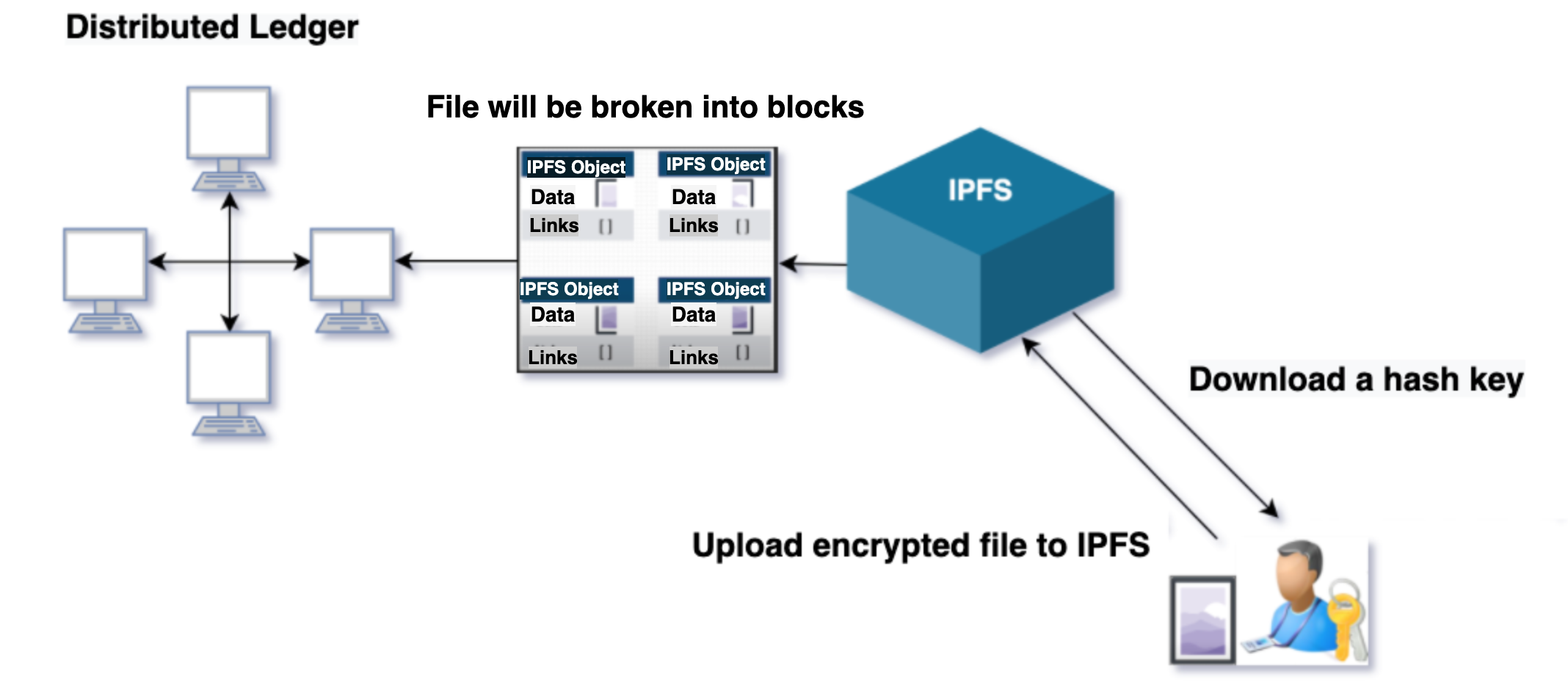}
\caption{Interplanetary file system, store and retrieve file.} 
\label{fig:Interplanetaryfilesystem}
\end{figure}

The InterPlanetary File System (IPFS) is an open-source, peer-to-peer network that is created as a decentralized system with the remarkable ability to store big data (scalability) and share files among distributed nodes globally, without the need of a central administrator \cite{politou2020delegated}. Therefore, it does not have a single point of failure. In the IPFS, the uploaded file is split into blocks and then distributed amongst the nodes, as shown in Figure~\ref {fig:Interplanetaryfilesystem} \cite{IPFS2021a}. Each block has a unique hash value representing the content identifier (CID) used to reference the file contents that need to be shared. Contrary to this, the traditional internet uses URL web addresses to share the objects based on the location approach. Many vital features contributed to emerging IPFS as a superior technique compared to others. These features result from the successful thinking of previous peer-to-peer systems, for instance, DHT, BitTorrent, Git, and SFS. In other words, IPFS is described as integrating solutions of prominent techniques and protocols in one platform. IPFS contains various techniques and protocols with different functionality, most prominent the CID, the DHT, and the Merkle DAG, Libp2p and the BitSwap, which represent essential pillars that IPFS is sitting on to attain the aim of file sharing and exchanging between scattered nodes globally. Overall, we conclude that IPFS combines various eminent technical concepts. Its mechanism involves several phases in which protocols and techniques work in harmony to share data between the IPFS network peers efficiently. IPFS achieves data storing, transferring, requesting, and deleting through key-value data storage management. On this basis, IPFS provides a high throughput, peers exchange, and share files in a content-based approach rather than a location-based approach, based on CID, Merkle DAG, DHT, Libp2p, and the BitSwap protocol that inspired from the Bit Torrent protocol \cite{henningsen2020mapping}.

\subsection{\large Hyperledger Fabric}

Hyperledger Fabric is a private-permissioned distributed ledger technology designed to be highly modular and extensible, delivering confidentiality, privacy and scalability as the de facto enterprise blockchain. Hyperledger Fabric does not involve any cryptocurrency as opposed to public blockchains; hence, it has a flexible design that gained a wide reputation in many different application areas \cite{papadopoulos2020privacy,stamatellis2020privacy}. It is provided as open-source by the Linux Foundation, and the smart contracts, namely chaincode, written in common programming languages such as Go, JavaScript, and Java, are being executed in a distributed operation manner \cite{androulaki2018hyperledger}. Hyperledger Fabric allows various components, such as complex consensus mechanisms and identity services, to be used as plug-and-play.

A \textit{smart contract} is a piece of programming code deployed in a blockchain environment that is being executed automatically if certain conditions are being met \cite{de2021smart}. Before accessing the blockchain ecosystem, all the participants have to agree on it, and \textit{install} it. Additionally, smart contracts involve the security policies and the blockchain rules set by its administrators \cite{wang2018blockchain, wood2021ethereum}. In Hyperledger Fabric, the smart contracts are called \textit{chaincode}, and are mutually accepted and installed in all the participating entities and define their permissions, the actions they can perform, and the access control policy to specific stored data \cite{androulaki2018hyperledger}.

\begin{figure}[ht!]
\centering
\includegraphics[width=1.05\linewidth]{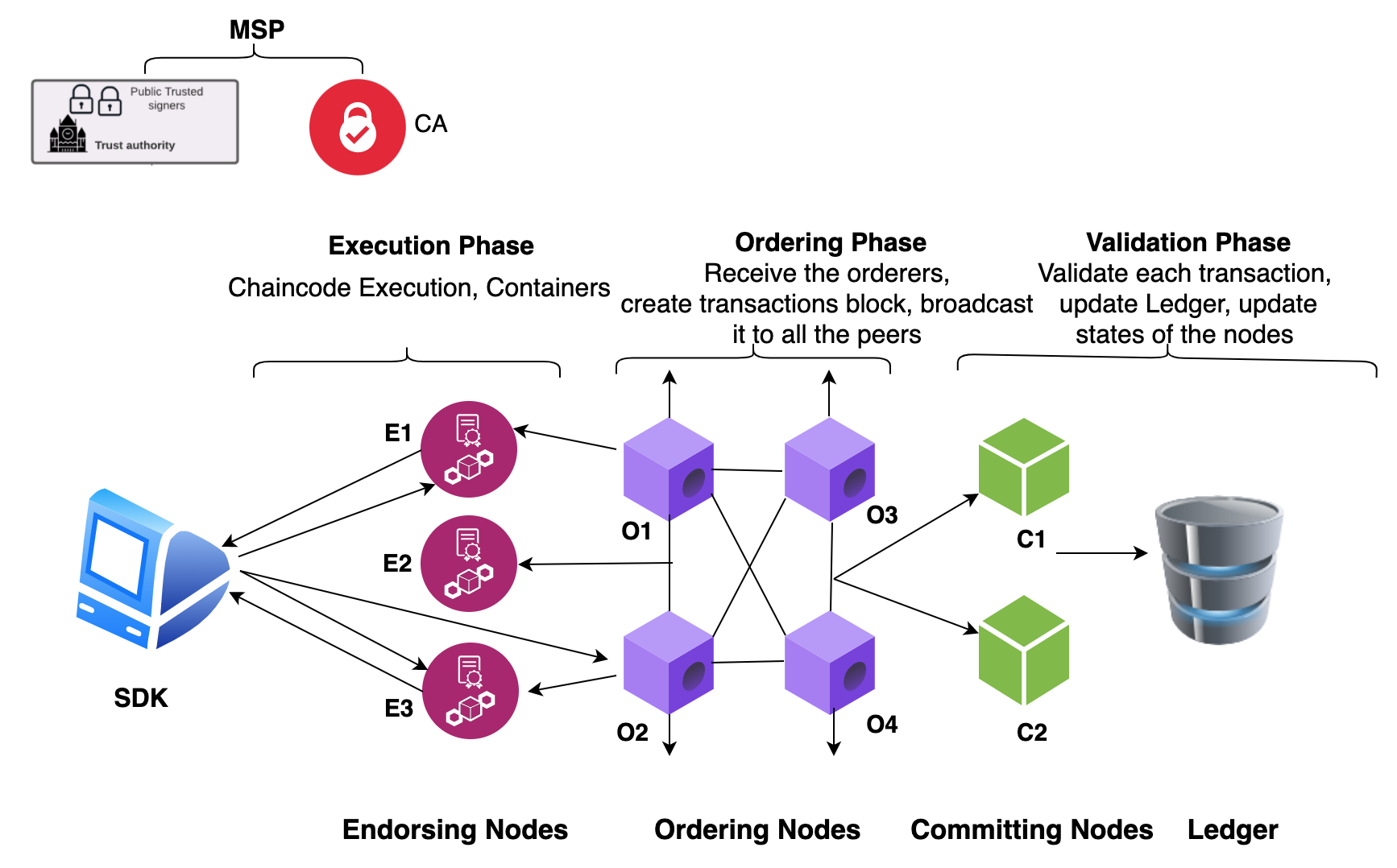}
\caption{Hyperledger Fabric operation phases}
\label{fig:HFL}
\end{figure}

\textit{Peers} is the fundamental entity participating in a Hyperledger Fabric ecosystem. The Peers can read from and write to the blockchain ledger according to their authorized identity certificates and defined permissions. Additionally, the Peers execute their installed chaincode to store their data to \textit{State Databases} such as LevelDB and CouchDB that store data in key-value pairs. State databases are used to track the global state of the blockchain ledger (the one that all the authorized participants have access to) and the private data collections, which are stored to private blockchain ledgers available only to authorized participants with the appropriate permissions. A group of Peers forms an \textit{Organization}, and multiple Organizations exist in a typical Hyperledger Fabric ecosystem. Multiple Organizations (and their Peers) can access specific blockchain ledgers that are available in their \textit{Channel}. Each Channel can have its own privacy rules and permissions; however, the Organizations can join multiple Channels using the same identity certificates. Within the Hyperledger Fabric community, it is considered a good practice to utilize separate Channels if certain participating Organizations do not need to ``share'' a blockchain ledger; however, in case that the Organizations may share some data stored in a blockchain ledger but not others, then \textit{Private Data Collections} can be used, allowing access only to specific data fields. Data can be stored in a blockchain ledger only after \textit{Consensus} between the Organizations is being met. This is one of the defined rules during the creation of the blockchain that specifies the number of Organizations required to approve a transaction to be considered valid. Hyperledger Fabric allows complex consensus mechanisms to be utilized, such as the Byzantine Fault-Tolerance. The \textit{Endorsement Policy} in Hyperledger Fabric defines which Peers of an Organization are required to execute the chaincode and approve or reject a transaction. These transaction proposals are forwarded to the \textit{Ordering Service} which checks if the number of Organizations that approve a transaction corresponds to the defined rules and, if successful, creates a new blockchain block to be added to the blockchain ledger and broadcasts it to all the associated Organizations and Peers in order to update their State Databases, as shown in Figure~\ref{fig:HFL} \cite{papadopoulos2020privacy,stamatellis2020privacy,androulaki2018hyperledger}.
 
In Hyperledger Fabric, the identity and permissions of each participant are defined through its identity certificates. By default, the Certificate Authority (CA) creates X.509 digital certificates for all the participants \cite{androulaki2018hyperledger}. These types of credentials follow a similar approach as the Public Key Infrastructure (PKI), in which the Hyperledger Fabric participants use their private keys to sign their transactions. In case that an X.509 certificate is being compromised, the CA can revoke its access to the blockchain ledgers and issue a new credential for its corresponding legitimate participant. The Membership Service Provider (MSP) verifies all the identity certificates for the participants in the ecosystem, as shown in Figure~\ref{fig:HFL}. This authentication and authorization process ensures that the stored data can be accessed only by legitimate participants with the appropriate permissions and not by unauthorized third parties. The usage of X.509 certificates can be further extended in more complex MSP implementations, and other types of digital identities can be used \cite{iftekhar2021hyperledger,papadopoulos2021privacy}.

\subsection{\large Decentralised Identities Management (DIDs)}

DIDs are a new type of digital identifier tool that provide systems trustworthy and privacy-preserving. The traditional identities are described as centralised management, which depends on the third party. Contrary, the DIDs are a distributed, decentralised self-authenticated system. DIDs are stored in a decentralised storage system such as IPFS and Distributed ledger as in the proposed framework. Furthermore, DIDs are managed and controlled by the organisation owner without third party need.

\begin{table*}[ht!]
\centering
\caption{Comparison of the related literature with our work.}
\label{tab:COMPARISON}
\begin{tabular}{|l|l|l|l|l|l|l|l|}
\hline
Related Work &
  Distributed Ledger &
  Accessibility &
  Verifiabilty &
  Scalability &
  GDPR &
  Content Erasure (RtbF) &
  Decentralised Identites \\ \hline
{\color[HTML]{000000} \cite{bawane2020ethegram}}   & Ethergram          & Yes & Yes & Yes, IPFS & Yes & No  & No  \\ \hline
{\color[HTML]{000000} \cite{havelange2019luce}}    & Blockchain Ledger  & Yes & Yes & No        & Yes & Yes & No  \\ \hline
{\color[HTML]{000000} \cite{wang2018data}} &
  Public Blockchain &
  Yes &
  Yes &
  Yes, Blockchain &
  Yes &
  No &
  No, Proxy Encryption \\ \hline
{\color[HTML]{000000} \cite{politou2020delegated}} & Blockchain         & Yes & Yes & Yes, IPFS & Yes & Yes & No  \\ \hline
{\color[HTML]{000000} \cite{grundstrom2019making}} &
  Hyperledger Fabric &
  Yes &
  Yes &
  Yes, IPFS &
  Yes &
  Yes &
  No \\ \hline
Our Work                                           & Hyperledger Fabric & Yes & Yes & Yes, IPFS & Yes & Yes & Yes \\ \hline
\end{tabular}
\end{table*}

\subsection{\large Related Works}

In the work of \cite{bawane2020ethegram}, the authors proposed the Ethegram framework to solve challenges of typical social media that derive from the centralization of systems. In contrast, the authors used the Ethereum blockchain and IPFS (InterPlanetary File System) peer-to-peer file system to distribute the stored content and address single point of failure issues. Additionally, the proposed decentralized system aims to solve censorship issues by using distributed peers without the need for a trusted intermediary. However, a challenge for the adoption of this architecture originates from the chosen blockchain since it involves a cryptocurrency; hence the transactions may be costly.

In the work of \cite{havelange2019luce}, the authors introduced the LUCE framework, which is a new method for data management, taking into account data accountability and their license terms. This framework utilizes a blockchain ledger to incentivize data sharing and reuse by facilitating compliance with licensing terms such as recording the data usage and its purpose. The essential purpose of creating the LUCE framework is to allow the individual data to be modified and deleted, complying with regulations such as the GDPR's rights to access, modification and deletion. This work is comparable with our proposed infrastructure in complying with the GDPR and licensing terms and incentivizing data sharing, but it does not consider the IPFS to distribute the data storage to address the scalability challenges.

In the work of \cite{wang2018data}, the authors proposed a novel infrastructure to solve data sharing issues concerning security, transparency, and traceability. The proposed solution suggests the use of two blockchains together. One blockchain is being used for the original data's storage and the other for the storage of the transactions. This infrastructure aims to share data securely and provide features such as linkability and traceability. In addition, this infrastructure is integrated with another proxy encryption technology to enhance the privacy and security of the system more. This article proposed dual-blockchain usage to solve the issues mentioned earlier; however, simply that is not sufficient to ultimately ensure the privacy of the stored data, especially since this system was developed with the scope to promote traceability of the stored data.

In the work of \cite{politou2020delegated}, the authors' combined blockchain technology and IPFS. They combined blockchain with the IPFS data sharing system to offer a broader data scale and store personal data off-chain. In other words, the original data is shared and distributed in the IPFS nodes rather than blockchain to avoid data censorship and comply with the \textit{Right to be Forgotten (RtbF)} of the GDPR. The European Union (GDPR) established the GDPR to enforce personal data protection within the EU, and all the associated entities in the same context \cite{grundstrom2019making}. Their proposed framework uses an anonymous protocol that aligns with the IPFS' principle to remove any Personally Identifiable Information (PII) from the data distributed over the IPFS nodes and ensures that an erasure request reaches all nodes using the delegated deletion mechanism. Besides that, only the owner or their delegates can erase the original content. The authors state that this mechanism for data erasure does not affect the system's overall efficiency; hence, it provides robust security with adequate performance for generating the content-dependent keys.

Table~\ref{tab:COMPARISON} shows how our work differentiates from the rest of previous work in the same context. It proposes a combination of a private-permissioned blockchain framework with IPFS to store threat information and share it amongst distributed peers. This combination ensures the stored content’s privacy, whereas it provides high performance in terms of network latency, transaction throughput, CPU and ram usage. Besides that, we use a decentralised identities management that ensures distribution identity in a secure, trusted and available manner.

MITRE ATT\&CK navigator is compatible with our proposed infrastructure. In which threat information consists of the tactics, techniques, and mitigation in the forms of IDs in the JSON file. It is presented in different colours and scoring. It will show the threat behaviour nicely, implicitly, simplistic, clear and comparable. The depictive of the threat scenario could be extracted in the JSON file format to be shared with the end-user. The MITRE ATT\&CK navigator represents the trusted partners' user interface, whereby threat sharing could be reviewed and contrasted to find the mitigation and best defence points. 

The partner will be identified by his node identity to download the sharing file on the local machine and then easy to upload, review, and contrast the threat dataset on the MITRE ATT\&CK navigator.


\section {\Large Methodology and Technical Architecture}

The proposed system consists of Hyperledger Fabric and Interplanetary File System (IPFS). This Infrastructure depicts the exact threat sharing mechanism using the MITRE ATT\&CK framework. Chaincode in the system are kept distributed because they are responsible for carrying out different organisations' policies and rules securely and privately.

\subsection {\large Data Sources}

Data sources represent the cumulative and real-life data associated with the case study that we need to analyse to find solutions for current and potential threats. Data sources consist of primary and secondary data. Primary data is collected from the original sources, reliable and contributes directly to the critical decision. The secondary data is collected from the literature review ( articles, journals, books, conferences, etc). Furthermore, data obtained from the expert was the most beneficial regarding the different experiments and comparisons in practical and theoretical aspects, avoided many complexities and efforts and saved searching time for valuable data.

Our measurements setup depends on a MacBook Pro with 1TB SSD drive, Memory 16 GB 1600 MHz DDR3, Processor 2.6 GHz Quad-Core Intel Core i7. In addition, we utilised the activity monitor application to measure the storage usage and processor utilisation. Our evaluation observed that our approach consumes low energy and CPU power.

\subsection {\large Data Analysis Framework}
 
 \begin{itemize}
 
 \item MITRE ATT\&CK navigator 

Most threat modelling and attack simulation works are done manually, which can be time-consuming and prone to errors. MITRE ATT\&CK is a globally accessible knowledge base of adversary tactics and techniques based on real-world observations\cite{holm2014automatic}. This knowledge base can be used as a foundation for developing specific threat models and other types of methodologies and tools. Our focus is on its navigator, the Detect Tactics, Techniques \& Combat Threats (DeTT\&CT) Editor.
 
The data logs (JSON file) will be imported to the MITRE ATT\&CK navigator to analyse and determine suitable defensive coverage. In contrast, we can export and share files representing the threats group of specific incidents to be viewed on the navigator back. In other words, the transformation process from the data logs and navigators is bidirectional. Therefore, this is considered beneficial concerning the data sharing and displaying process and getting a trusted and secure way between the scatted partners.

MITRE ATT\&CK navigator has other remarkable features that ease the security analysts, such as colouring comparing, scoring and combining different groups of the threat actions in a clear view \cite{2015-2021MITREATTCK}.

\item DeTT\&CT Editor 
In April 2019, Ruben and Marcus Bakker released the first version of DeTT\&CT. It was created at the Cyber Defence Centre of Rabobank and built on top of the MITRE ATT\&CK \cite{DeTTCTsGNUGeneralPublicLicensev3.0}. 

The DeTT\&CT Editor has been created to assist blue teams using ATT\&CK to score and compare data log source quality, visibility coverage, detection coverage and threat actor behaviours. All of which can help, in different ways, to get more resilient against attacks targeting organisations.

\end{itemize}
 
\subsection {\large Use Case: Threats Information Sharing}
 
The use cases were obtained from the MITRE ATT\&CK framework related to the previous real-world threats experiments of industry and manufacturing fields subjected to cyber security threats and interested in threat intelligence sharing. We will describe and analyse some cyberattack incidents identified in healthcare, such as WICKED PANDA and Fox Kitten.

The WICKED PANDA threat group is identified by "ID: G0096" and associated Groups called WICKED PANDA, an espionage group that conducts financially motivated operations. The group has been observed targeting healthcare, telecom, technology, and video game industries in 14 countries. We will take this case study to demonstrate how to use the ATT\&CK technique methodology to determine when and how to develop new techniques for ATT\&CK for ICS \cite{APT41:WICKEDPANDA}. 

\begin{figure}[ht!]
\centering
\includegraphics[width=0.757\linewidth]{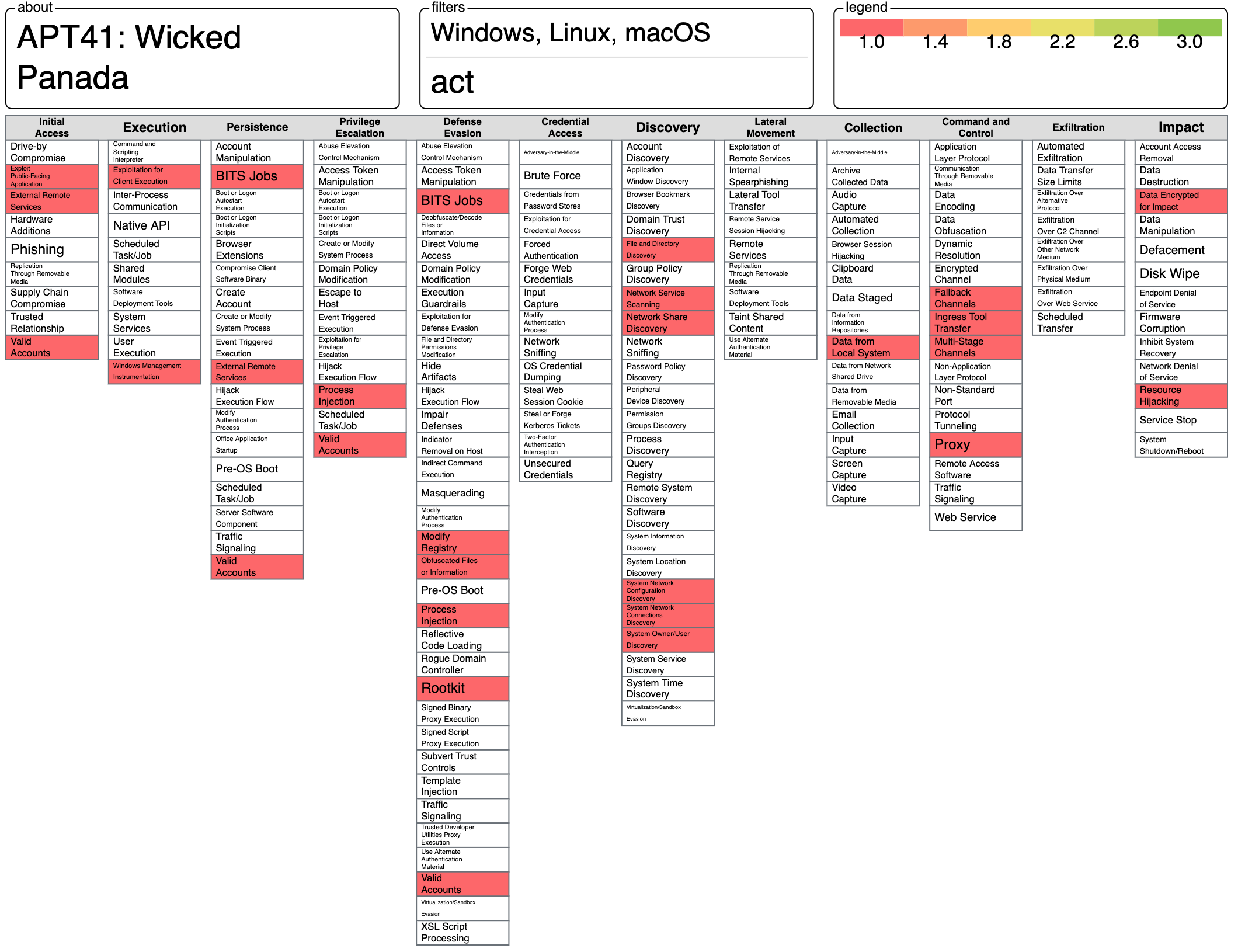}
\caption{WICKED PANDA Threat, Attacker Tactics and Techniques}
\label{fig:WICKEDPANDA}
\end{figure}

Fox Kitten threat group targets multiple industries, including oil and gas, technology, government, defence, healthcare, manufacturing, and engineering. It is associated with areas and companies such as UNC757, PIONEER KITTEN, Parisite, the Middle East, North Africa, Europe, Australia, and North America \cite{FoxKitten}.

\begin{figure}[ht!]
\centering
\includegraphics[width=0.757\linewidth]{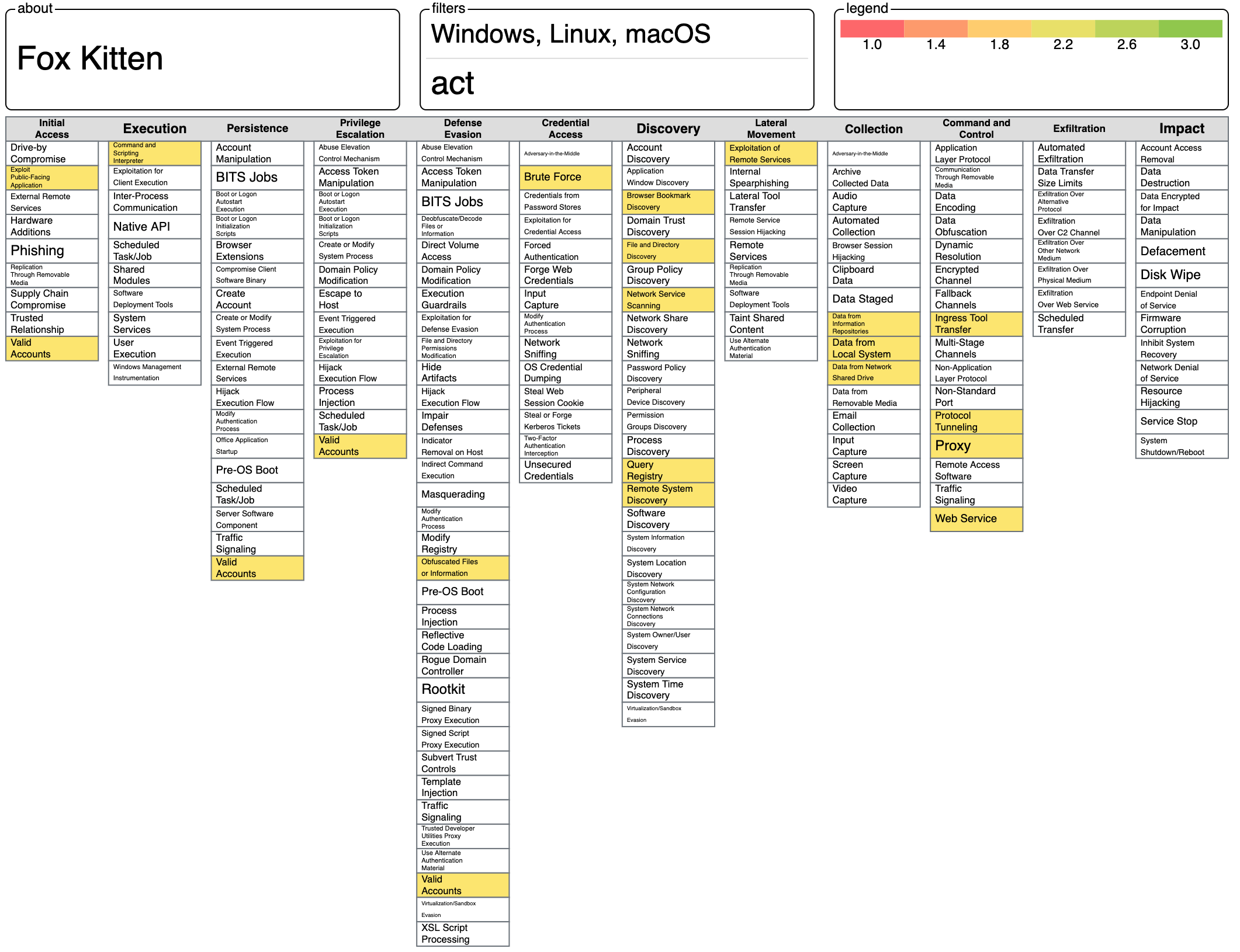}
\caption{Fox Kitten Threat, Attacker Tactics and Techniques}
\label{fig:FoxKitten}
\end{figure}

\begin{figure}[ht!]
\centering
\includegraphics[width=0.77\linewidth]{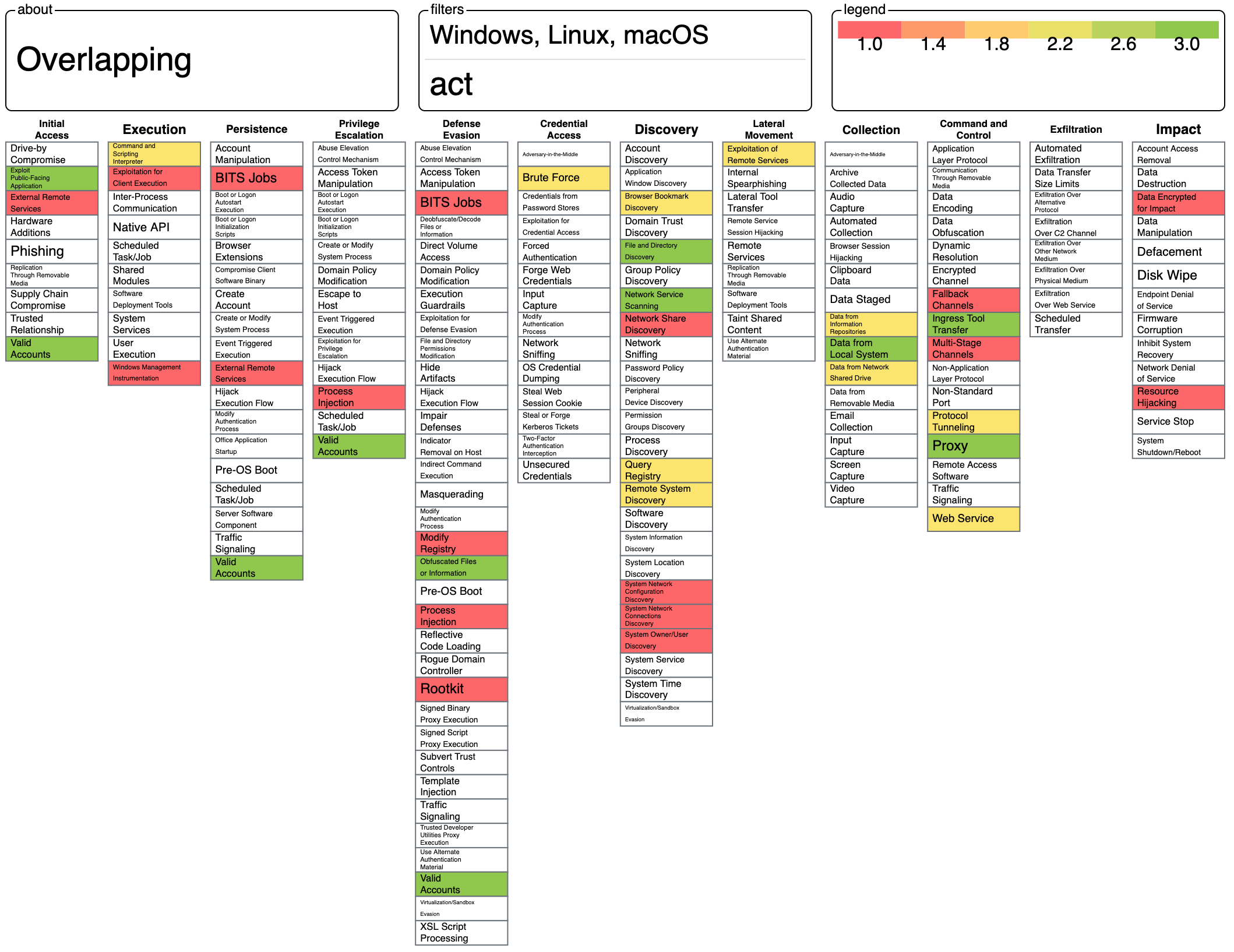}
\caption{Overlapping between Wicked Panda layer and Fox Kitten layer illustrated by the green colour}
\label{fig:overlapping}
\end{figure}

\subsection {\large The Proposed Method and Collaborative Operations using IPFS and Hyperledger Fabric}
 
Our proposed ecosystem provides mechanisms to share sensitive data and private threat information across broad IT and public networks in a secure and trusted way. We have chosen an eminent use case of real-world observations of the attacker's behaviours and systems breach in the healthcare domain. Threats information obtained in JSON files, viewed and contrasted using MITRE ATT\&CK navigation as shown in Figures \ref{fig:mitrenavigator}, \ref{fig:WICKEDPANDA}, \ref{fig:FoxKitten}, and \ref{fig:overlapping}.

The function of each operation number and flow of the system is illustrated in Figure~\ref{fig:mitresharing}, and explained, as follows:

\begin{enumerate}

  \item New threat is stored in a local database system.
  
  \item Threat files of WICKED PANDA and Fox Kitten are uploaded to ATT\&CK navigator to analyse and compare the adversary behaviour and find suitable countermeasures.
  
  \item The MSP (DIDs) provides nodes identities and certificates to authenticate the interacted parties. The sender encrypts and signs the threat file (JSON file).
  
   \item Sender uploads the encrypted file to IPFS. 
  
  \item Sender downloads the hash file (CID) from IPFS. 
 
  \item  Sender submits the transaction (CID) using its own identity. Sender is authenticated by chaincode, endorsement. 
  \item  The trusted partner is authenticated by chaincode and then get the CID using his private key. 
  
  \item A trusted partner will use CID to search and get access to the threat file into IPFS storage.
   
  \item A trusted partner will download the threat desired file from IPFS using the CID.
  
  \item The trusted partner retrieves threat file from IPFS, decrypt it and then display the WICKED PANDA and Fox Kitten threats on the MITRE ATT\&CK navigation. 
 
\end{enumerate}

\begin{figure}[ht!]
\centering
\includegraphics[width=1\linewidth]{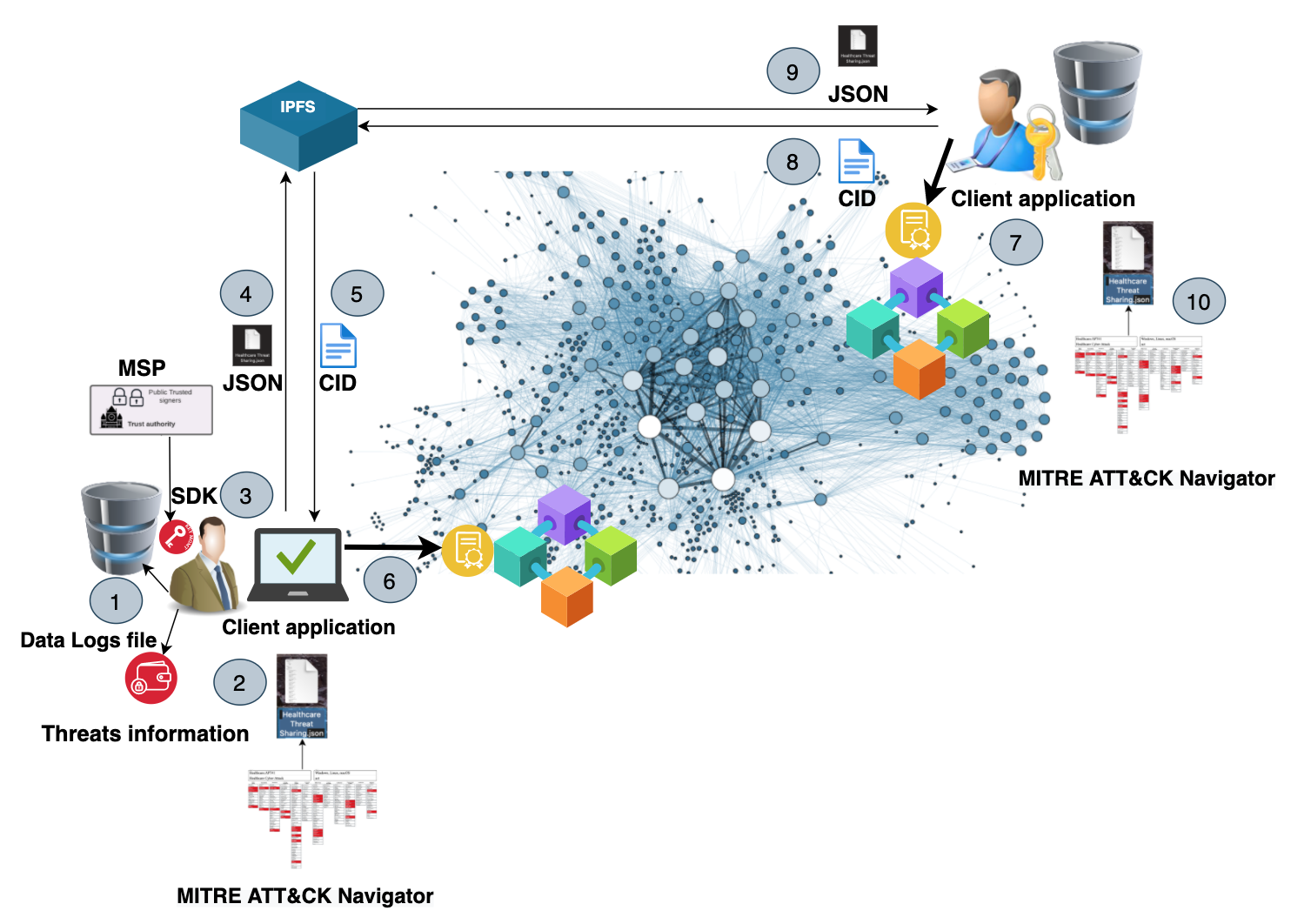}
\caption{Threat information flow in proposed infrastructure}
\label{figure:Threatinformation }
\end{figure}

As a result, the proposed ecosystem can share and store threat files in a secure and trusted way. The combination of Hyperledger Fabric and InterPlanetary File System (IPFS) works together highly efficiently. Hyperledger Fabric, InterPlanetary File System (IPFS), DIDs and MITRE ATT\&CK framework guarantees privacy-preserving and threat sharing, securely plus scalability and anonymity. 

\section{\large Empirical Evaluation}

The key objective of our paper is to implement a data-sharing infrastructure within a permissioned ledger, which creates an ecosystem for the trusted sharing of threat information. To do so, we consider the system's security and anonymity, latency, and scalability.

\subsection{\large Security Consideration}

From a security point of view, our proposed infrastructure is a decentralised network that is resistant to denial of service attacks since it does not have a single administrator service, and its data content is distributed among the participating nodes. Besides that, Hyperledger Fabric supports transaction encryption and chaincode that governs the interaction between the trusted users, ensuring transaction authentication and data confidentiality. Any user should pass an identity check; otherwise, their transaction will be declined.  

Furthermore, we use the MSP entity, which is considered the key element that offers the system authentication, as a part of the Hyperledger Fabric that contains certificate authority and Identities management. This entity supplies all the nodes and channels with unique identities and enables transactions to be private and secure when traversing the network between the interacting users. By default, the system is protected against unauthorised users. A potential vulnerability to our proposed system could be the human factor since their credentials may be compromised intentionally or negligently. 

In light of the above, the certificate authority is the third entity that supports the interacting parties to trust each other using its signature as the proof of identity. If any interacting parties are compromised, they can ask the certificate authority to revoke their transactions. As a result, any adversary targeting the Hyperledger Fabric by establishing read or write queries will be exposed since it does not own any system identities. The only way is to compromise an authorised user's identity. Another potential threat is that adversaries may target the blockchain ledgers since they are disseminated across the participating nodes, and the legitimate users' access to them may be blocked.

\subsection{\large Anonymity consideration}

The Hyperledger Fabric supports transaction anonymity by default since it includes the MSP entity that is responsible for identity management. On top of that, this unit is pluggable, meaning that each organization could create its own MSP.

Data transaction anonymity derives from the transaction data encryption, by the client's private key and signed by the trusted third party (identity management), and the Transport Layer Security cryptographic protocol (TLS) that is used to encrypt the communication between the interacting participants. The authorized user will be identified in the chaincode and should gather enough of the endorsements to be able to send the transaction for validation, which is governed by the endorsers' policy without disclosing or revealing the transaction data.  

In the work of \cite{mazumdar2019design}, the authors propose an Anonymous Endorsement System which uses a threshold endorsement policy. Several factors contribute to the design of a new ring signature scheme, called Fabric' Constant-Sized Linkable Ring Signature (FCsLRS) with Transaction-Oriented linkability for hiding the identity of the endorsers. Additionally, this work implemented the signature scheme in Golang and analyzed its security and performance by varying the RSA (Rivest-Shamir-Adleman) modulus size. The feasibility of the implementation is supported by empirical analysis. The signature and tag generation time is relatively fast. Moreover, it remains constant irrespective of change in message length or endorsement set size for a given RSA modulus value, assuming all the endorsers generate their signatures in parallel.

\subsection{\large Latency, Scalability, Throughput}

Since Hyperledger Fabric is a permissioned-private blockchain framework, the read and write queries' latency is comparable to common centralized systems, such as typical databases (PostgreSQL), instead of public blockchain systems such as Bitcoin. However, it should be noted that in a higher number of transaction records, our chosen infrastructure performs more efficiently contrary to typical local databases, with the transaction times being reduced \cite{papadopoulos2020privacy,stamatellis2020privacy}.

In terms of scalability, since the Hyperledger Fabric is utilizing Docker containers to act as nodes of the blockchain, it is easily scalable to cloud infrastructures such as Kubernetes clusters. This adds immense extensibility to the given use cases by incorporating several Peers and Organizations to create complex scenarios. Additionally, in our proposed ecosystem, the actual records can be stored distributed in IPFS; hence, extending the scenario even more.

The transaction throughput of our system depends on the ordering service and is capped by the network and ordering nodes capacity size. However, simply by adding a higher number of ordering nodes, the transaction throughput can be increased. Examples of adding more nodes include RAFT and KAFKA ordering clusters \cite{papadopoulos2020privacy}.

 
\section{ Conclusion and Future Work}

Threat hunting is fundamentally a proactive countermeasure for detecting and protecting IT systems against malicious behaviour by monitoring current and historical information related to security breaches. Therefore, it is necessary to come up with a trusted and secure framework for threat information sharing. The deployment of trusted computing technologies for adversary detection and privacy protection can help achieve this goal using data logs generated by systems, network devices or security applications such as intrusion detection/prevention systems. Our work proposed a trusted information sharing solution utilizing both the IPFS system and a permissioned ledger that ensures the security, privacy, and anonymity of the stored information that achieves high throughput with immense scalability. The permisioned distributed ledger technology of our choice is the Hyperledger Fabric, which meets all the aforementioned expectations and achieves our work's aims. 

Our proposed infrastructure implements threat information sharing effectively, using the MITRE ATT\&CK Framework, the pluggable certificate authorities, and the self-executing chaincode when a set of defined conditions is being met, thus enabling the establishment of trust between the interacting parties and enhancing the total security of the system. Hence, our ecosystem ensures the anonymity of the stored information. As future work, we aim to build a comprehensive proof-of-concept in a cloud infrastructure, utilizing a Kubernetes cluster to improve the system's throughput and scalability considering the expansion of the participating nodes and the transactions' capacity within the GLASS project, which represents the ongoing European Union-funded project in this context.

\section*{Acknowledgements}
The research leading to these results has been partially funded by the European Union's Horizon 2020 research and innovation programme, through funding of the GLASS project (under grant agreement No 959879).

\bibliographystyle{IEEEtran}
\bibliography{ref}

\end{document}